\pdfoutput=1

\def\year{2019}\relax
\documentclass[letterpaper]{article}
\usepackage{aaai19}
\usepackage{times}
\usepackage{helvet}
\usepackage{courier}
\usepackage[hyphens]{url}
\usepackage{graphicx}
\usepackage{graphicx}
\usepackage{fancyvrb}
\usepackage{cprotect}
\usepackage{xspace}
\usepackage{booktabs}
\usepackage{array}
\usepackage{multirow}
\usepackage{makecell}

\newcommand*{\wlg}{\textsc{WikiLinkGraphs}\xspace}
\newcommand*{\wikilink}{\emph{wikilink}\xspace}

\nocopyright

\title{WikiLinkGraphs: A Complete, Longitudinal and Multi-Language Dataset of the Wikipedia Link Networks}

\author{Cristian Consonni\\
{DISI, University of Trento}\\
cristian.consonni@unitn.it
\And
David Laniado\\
{Eurecat, Centre Tecnol\`ogic de Catalunya}\\
david.laniado@eurecat.org
\And
Alberto Montresor\\
{DISI, University of Trento}\\
alberto.montresor@unitn.it}

\begin{document}

\maketitle

\begin{abstract}
\noindent Wikipedia articles contain multiple links connecting a subject to other pages of the encyclopedia. In Wikipedia parlance, these links are called internal links or \mbox{\emph{wikilinks}}. We present a complete dataset of the network of internal Wikipedia links for the $9$ largest language editions. The dataset contains yearly snapshots of the network and spans $17$ years, from the creation of Wikipedia in 2001 to March 1st, 2018.
While previous work has mostly focused on the complete hyperlink graph which includes also links automatically generated by templates, we parsed each revision of each article to track links appearing in the main text. 
In this way we obtained a cleaner network, discarding more than half of the links and representing all and only the links intentionally added by editors.
We describe in detail how the Wikipedia dumps have been processed and the challenges we have encountered, including the need to handle special pages such as \emph{redirects}, i.e., alternative article titles.
We present descriptive statistics of several snapshots of this network. 
Finally, we propose several research opportunities that can be explored using this new dataset.
\end{abstract}

\section{Introduction}
Wikipedia\footnote{\url{https://www.wikipedia.org}} is probably the largest existing information repository, built by thousands of volunteers who edit its articles from all around the globe. As of March 2019, it is the fifth most visited website in the world~\cite{alexa:topsites}. Almost 300k active users per month contribute to the project~\cite{wiki:en:List_of_Wikipedias}, and more than 2.5 billion edits have been made. The English version alone has more than 5.7 million articles and 46 million pages and is edited on average by more than 128k active users every month~\cite{wiki:en:Special:Statistics}.
Wikipedia is usually a top search-result from search engines~\cite{lewandowski2011ranking} and research has shown that it is a first-stop source  for information of all kinds, including information about science~\cite{yasseri2014most,spoerri2007popular}, and medicine~\cite{laurent2009seeking}.

The value of Wikipedia does not only reside in its articles as separated pieces of knowledge, but also in the links between them, which represent connections between concepts and result in a huge conceptual network.
According to Wikipedia policies\footnote{In what follows, we will refer to the policies in force on the English-language edition of Wikipedia; we will point out differences with local policies whenever they are relevant.}~\cite{wiki:en:Wikipedia:MOS:UL}, when a concept is relevant within an article, the article should include a link to the page corresponding to such concept~\cite{borra2015societal}. Therefore, the network between articles may be seen as a giant mind map, emerging from the links established by the community. Such graph is not static but is continuously growing and evolving, reflecting the endless collaborative process behind it.

The English Wikipedia includes over 163 million connections between its articles. This huge graph has been exploited for many purposes, from natural language processing~\cite{yeh2009wikiwalk} to artificial intelligence~\cite{navigli2012babelnet}, from Semantic Web technologies and knowledge bases~\cite{presutti2014uncovering} to complex networks~\cite{capocci2006preferential}, from controversy mapping~\cite{markusson2016contrasting} to human way-finding in information networks~\cite{west2012human}.

This paper presents a new dataset, \wlg, that makes the networks of internal links in the nine largest editions of Wikipedia available to researchers and editors, opening new opportunities for research. 
Most previous work on the Wikipedia link graph relies on wikilink data made accessible through the Wikipedia API\footnote{Hyperlinks in the current version of Wikipedia are available through the "Link" property in the Wikipedia API: \url{https://www.mediawiki.org/wiki/API} } and through database dumps\footnote{\url{https://meta.wikimedia.org/wiki/Data_dumps}}
These data include also all transcluded links, i.e. links automatically generated by templates defined in another page; templates typically add all possible links within a given group of articles, producing big cliques and inflating the density of connections.

\begin{figure*}[t!]
\centering
\includegraphics[width=0.95\textwidth]{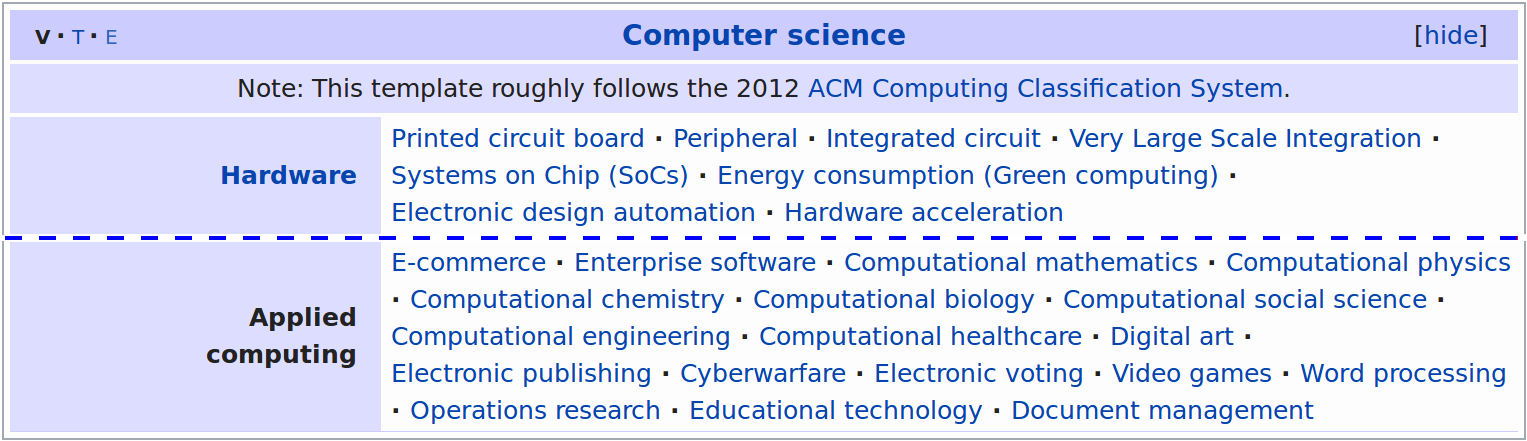}
\caption{\label{image:template} A portion of the navigational template \texttt{\{\{Computer science\}\}} from English Wikipedia as of revision n\textsuperscript{o} 878025472 of 12 January 2019, 14:12. The dashed line indicates that a portion of template has been stripped for reasons of space.}
\end{figure*}

Inserting a template in Wikipedia merely amounts to writing a small snippet of code, which in the final article is rendered as a collection of links. Figure~\ref{image:template} shows a rendering of the navigation template \texttt{\{\{Computer science\}\}}\footnote{\url{https://en.wikipedia.org/wiki/Template:Computer_science}} from English Wikipedia, which produces a table with 146 links to other articles within the encyclopedia. Navigation templates are very general by design serve to group links to multiple related articles. They are not specific to a given page: in fact, the content of a template can be changed independently from editing the pages where it is included.

We argue that considering only the links explicitly added by editors in the text of the articles may provide a more trustful representation of semantic relations between concepts, and result in a cleaner graph by avoiding the cliques and other potential anomalous patterns generated by transcluded links.

The aim of this work is to build a dataset of the graph of the specific link between Wikipedia articles added by the editors. The \wlg dataset was created by parsing each article to extract its links, leaving only the links intentionally added by editors; in this way, we discarded over half of the overall links appearing in the rendered version of the Wikipedia page.

Furthermore, we tracked the complete history of each article and of each link within it, and generated a dynamic graph representing the evolution of the network. Whilst the dataset we are presenting in this paper consists of yearly snapshots, we have generated several supporting dataset as well, such as a large collection tracking the timestamp in which each occurrence of a link was created or removed.

Redirects, i.e. special pages representing an alternative title for an article, are a known issue that was shown to affect previous research~\cite{hill2014consider}. In our dataset, we tracked the redirects over time, and resolved all of them according to the corresponding timestamp. The complete history of all redirects is made available as well.

The code used to generate the dataset is also entirely made available on GitHub, so that anybody can replicate the process and compute the wikilink graphs for other language editions and for future versions of Wikipedia. 

\section{The \textsc{WikiLinkGraphs} Dataset}

This section describes how we processed the Wikipedia dumps of the complete edit history to obtain the dataset.

\subsection{Data Processing}

The \wlg dataset was created from the full Wikipedia revision history data dumps of March 1, 2018\footnote{All files under "All pages with complete edit history (.7z)" at \url{https://dumps.wikimedia.org/enwiki/20180301/}. Wikipedia dumps are available up to 3 months prior to the current date, so those specific dumps are not available anymore. However, any dump contains the whole Wikipedia history dating from 2001 onwards. So our results can be replicated with any dump taken later than March 1st, 2018.}, as published by the Wikimedia Foundation, and hence includes all entire months from January 2001 to February 2018.

These XML dumps contain the full content of each Wikipedia page for a given language edition, including encyclopedia articles, talk pages and help pages. Pages are divided in different \emph{namespaces}, that can be recognized by the prefix appearing in the title of the page. The encyclopedia articles are in the \emph{main namespace}, also called \texttt{namespace 0} or \texttt{ns0}. The content of the pages in Wikipedia is formatted with \emph{Wikitext}~\cite{wiki:en:Help:Wikitext}, a simplified syntax that is then rendered as HTML by the MediaWiki software\footnote{\url{https://www.mediawiki.org}}. For each edit a new \emph{revision} is created: the dump contains all revisions for all pages that were not deleted.

\begin{table}[ht]
\centering
\begin{tabular}{lrrrr}
\toprule
\multicolumn{1}{l}{\textbf{lang}} & %
\multicolumn{1}{c}{\textbf{\makecell{size\\(GB)}}} & %
\multicolumn{1}{c}{\textbf{files}} & %
\multicolumn{1}{c}{\textbf{pages}} & %
\multicolumn{1}{c}{\textbf{revisions}} %
\\ %
\midrule
de  & 33.0    & 109  & 3,601,030   & 113,836,228    \\
en  & 138.0   & 520  & 13,750,758  & 543,746,894    \\
es  & 27.0    & 68   & 3,064,393   & 77,498,219     \\
fr  & 26.0    & 95   & 3,445,121   & 99,434,840     \\
it${}^{\dagger}$  & 91.0    & 61   & 2,141,524   & 68,567,721     \\
nl  & 7.4     & 34   & 2,627,328   & 38,226,053     \\
pl  & 15.0    & 34   & 1,685,796   & 38,906,341     \\
ru  & 24.0    & 56   & 3,362,946   & 63,974,775     \\
sv  & 9.0     & 1    & 6,139,194   & 35,035,976     \\ \bottomrule
\end{tabular}
\caption{\label{table:dumps} Statistics about the processed Wikipedia dumps:
size of the dowloaded files and number of processed pages and revisions for each dump.
(${}^{\dagger}$) the Italian Wikipedia dumps were downloaded in \texttt{.bz2} format.}
\end{table} 
Table~\ref{table:dumps} presents the compressed sizes for the XML dumps that have been downloaded and the number of pages and revisions that have been processed. We extracted all the article pages. This resulted in 40M articles being analyzed. In total, more than 1B revisions have been processed to produce the \mbox{\textsc{WikiLinkGraphs}}\xspace dataset.

\subsubsection{Link Extraction}

Wikipedia articles have \emph{revisions}, which represent versions of the Wikitext of the article at a specific time.
Each modification of the page (an \emph{edit} in Wikipedia parlance) generates a new revision. Edits can be made by \emph{anonymous} or \emph{registered} users.

A revision contains the wikitext of the article, which can have sections, i.e. header titles. Sections are internally numbered by the MediaWiki software from $0$, the \emph{incipit} section, onwards. As for HTML headers, several section levels are available (sections, subsections, etc.); section numbering does not distinguish between the different levels.

While a new visual, WYSIWYG editor has been made available in most Wikipedia editions starting since June 2013~\cite{wiki:en:Wikipedia:VisualEditor}, the text of Wikipedia pages is saved as \emph{Wikitext}. In this simplified markup language, internal Wikipedia links have the following format {\verb![[title|anchor]]!}; for example,

\begin{figure}[h!]
\centering
\begin{BVerbatim}
[[New York City|The Big Apple]]
\end{BVerbatim}
\end{figure}

This wikitext is visualized as the words {\verb!The Big Apple!} that gets translated into HTML as: 

\begin{figure}[h!]
\centering
\begin{BVerbatim}[fontsize=\small]
<a href="/wiki/New_York_City"
   title="New York City">The Big Apple</a>
\end{BVerbatim}
\end{figure}

\noindent pointing to the Wikipedia article \emph{New York City}. If the page exists, as in this example, the link will be blue-colored, otherwise it will be colored in red, indicating that the linked-to page does not exist~\cite{wiki:en:Wikipedia:Red_link}. The anchor is optional and, if it was omitted, then the page title, in this case {\verb!New York City!}, would have been visualized.

For each revision of each page in the Wikipedia dump, we used the following regular expression in Python\footnote{{\scriptsize \url{https://github.com/WikiLinkGraphs/wikidump/blob/70b0c7f929fa9d66a220caf11c9e31691543d73f/wikidump/extractors/misc.py#L203}}} to extract \emph{wilinks}:

\begin{Verbatim}[numbers=left,xleftmargin=5mm]
\[\[
(?P<link>
   [^\n\|\]\[\<\>\{\}]{0,256}
)
(?:
  \|
  (?P<anchor>
      [^\[]*?
  )
)?
\]\]
\end{Verbatim}
 
Line~1 matches two open brackets; then, Lines~2--4 capture the following characters in a named group called \verb!link!. 
Lines~5--10 match the optional anchor: Line~5 matches a pipe character, then Lines~6--8 match non-greedily any valid character for an anchor saving them in a named group called \verb!anchor!. Finally, Line~10 matches two closed brackets. The case of links pointing to a section of the article is handled \emph{a posteriori}, after the regular expression has captured its contents. When linking to a section, the \texttt{link} text will contain a pound sign (\verb!#!); given that this symbol is not allowed in page titles, we can separate the title of the linked page from the section.

\paragraph{The \textsc{RawWikilinks} Dataset.}

\begin{figure*}[h!]
\centering
\includegraphics[width=0.95\textwidth]{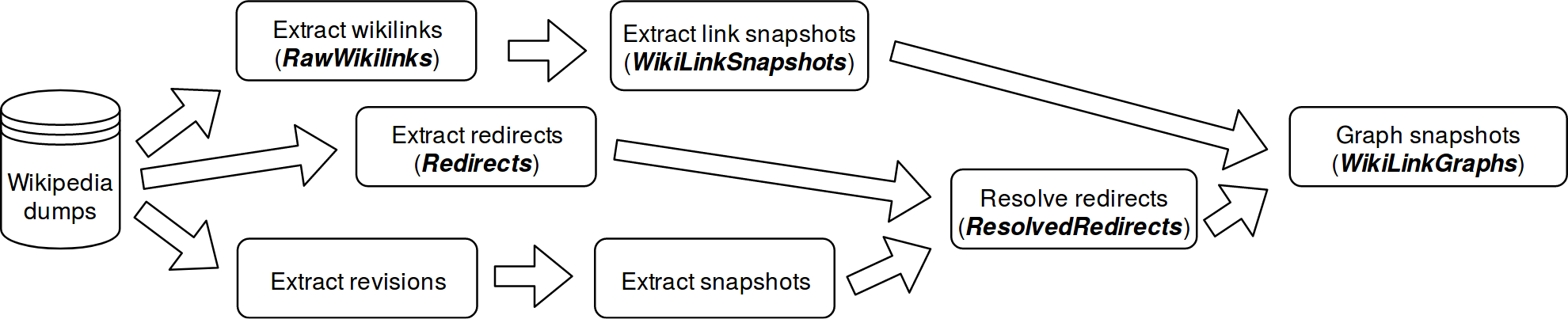}
\caption{\label{image:process} The process to produce the \wlg dataset from the Wikipedia dumps. In bold and italics the name of the intermediate datasets produced.}
\end{figure*}

The link extraction process produces a dataset with the following information:
\begin{itemize}
  \item \verb!page_id!: an integer, the page identifier used by MediaWiki. This identifier is not necessarily progressive, there may be gaps in the enumeration;
  \item \verb!page_title!: a string, the title of the Wikipedia article;
  \item \verb!revision_id!: an integer, the identifier of a revision of the article, also called a \emph{permanent id}, because it can be used to link to that specific revision of a Wikipedia article;
  \item \verb!revision_parent_id!: an integer, the identifier of the parent revision. In general, each revision as a unique parent; going back in time before 2002, however, we can see that the oldest articles present non-linear edit histories. This is a consequence of the import process from the software previously used to power Wikipedia, MoinMoin, to MediaWiki;
  \item \verb!revision_timestamp!: date and time of the edit that generated the revision under consideration;
  \item \verb!user_type!: a string ("\texttt{registered}" or "\texttt{anonymous}"), specifying whether the user making the revision was logged-in or not;
  \item \verb!user_username!: a string, the username of the user that made the edit that generated the revision under consideration;
  \item \verb!user_id!: an integer, the identifier of the  user that made the edit that generated the revision under consideration;
  \item \verb!revision_minor!: a boolean flag, with value 1 if the edit that generated the current revision was marked as \emph{minor} by the user, 0 otherwise;
  \item \verb!wikilink.link!: a string, the page linked by the \\wikilink;
  \item  \verb!wikilink.tosection!: a string, the name of the section if the link points to a section;
  \item \verb!wikilink.anchor!: a string, the anchor text of the \wikilink;
  \item \verb!wikilink.section_name!: the name of the section wherein the \wikilink appears;
  \item \verb!wikilink.section_level!: the level of the section wherein the \wikilink appears;
  \item \verb!wikilink.section_number!: the number of the section wherein the
  \wikilink appears.
\end{itemize}

\subsubsection{Redirects and Link Resolution}

A redirect in MediaWiki is a page that automatically sends users to another page. For example, when clicking on a \wikilink {\verb![[NYC]]!}, the user is taken to the article \emph{New York City} with a note at the top of the page saying: "(Redirected from NYC)". The page \emph{NYC}\footnote{\url{https://en.wikipedia.org/w/index.php?title=NYC&redirect=no}} contains special Wikitext:
\begin{BVerbatim}
#REDIRECT [[New York City]]
\end{BVerbatim}
\noindent which defines it as a redirect page and indicates the target article. It is also possible to redirect to a specific section of the target page. Different language editions of Wikipedia use different words\footnote{{\scriptsize \url{https://github.com/WikiLinkGraphs/wikidump/blob/70b0c7f929fa9d66a220caf11c9e31691543d73f/wikidump/extractors/redirect.py#L14}}}, which are listed in Table~\ref{table:redirects}.

\begin{table}[ht]
\centering
\begin{tabular}{ll}
\toprule
\multicolumn{1}{c}{\textbf{lang}}     & \textbf{words}                                        \\ \midrule
de                & {\verb!#WEITERLEITUNG!}                               \\
en                & {\verb!#REDIRECT!}                                    \\
es                & {\texttt{\#REDIRECCI\'ON}}, {\verb!#REDIRECCION!}          \\
fr                & {\verb!#REDIRECTION!}                                 \\
it                & {\verb!#RINVIA!}, {\verb!#RINVIO!}, {\verb!#RIMANDO!} \\
nl                & {\verb!#DOORVERWIJZING!}                              \\
pl                & {\verb!#PATRZ!}, {\verb!#PRZEKIERUJ!}, {\verb!#TAM!}  \\
ru${}^{\ddagger}$ & {\verb!#PERENAPRAVLENIE!}, {\verb!#PERENAPR!}         \\
sv                & {\verb!#OMDIRIGERING!}                                \\ \bottomrule
\end{tabular}
\cprotect\caption{\label{table:redirects} Words creating a redirect in MediaWiki for different languages. {\verb|#REDIRECT|} is valid on all languages. (${}^{\ddag}$) For Russian Wikipedia, we present the transliterated words.}
\end{table}
 
In general, a redirect page can point to another redirect page creating a chain of multiple redirects\footnote{For example, a live list of pages creating chains of redirect on English Wikipedia is available at \url{https://en.wikipedia.org/wiki/Special:DoubleRedirects}.}. These pages should only be temporary and they are actively eliminated by Wikipedia volunteers manually and using automatic scripts.

Despite the name, redirects are served as regular pages by the MediaWiki software so requesting a redirect page, for example by visiting the link \url{https://en.wikipedia.org/wiki/NYC}, returns an HTTP status code of 200.

\subsubsection{Resolving Redirects}\label{sss:resolving_redirects}

We have extracted one snapshot per year on March, 1st from the \textsc{RawWikilinks} dataset. The creation of a snapshot for a given year entails the following process:
\begin{enumerate}
\item we list all \emph{revisions} with their timestamps from the dumps;
\item we filter the list of revisions keeping only those that existed on March 1st, i.e. the last revision for each page created before March 1st;
\item we resolve the redirects by comparing each page with the list of redirects obtained as described above;
\end{enumerate}

At the end of this process, we obtain a list of the pages that existed in Wikipedia on March, 1st of each year, together with their target, if they are redirects. We call this dataset \textsc{ResolvedRedirects}.

It should be noted that even if we resolve redirects, we do not eliminate the corresponding pages: in fact, redirects are still valid pages belonging to the \texttt{namespace 0} and thus they still appear in our snapshots as nodes with one outgoing link, and no incoming links.

\subsubsection{Link Snapshots}

We then process the \textsc{RawWikilinks} dataset and we are able, for each revision of each page, to establish whether a wikilink in a page was pointing to an existing page or not. 
We add this characteristics to the \textsc{RawWikilinks} dataset in the field \verb!wikilink.is_active!: a boolean representing whether the page pointed to by the link was existing in that moment or not. Revisions are then filtered so to obtain the lists of links existing in each page at the moment of interest; we call this new dataset \textsc{WikiLinkSnapshots}.

\subsubsection{Graph Snapshots (\wlg)}\label{sss:graph_snapshots}

Armed with the \textsc{WikiLinkSnapshots} and the \textsc{ResolvedRedirects} dataset we can extract the \wlg as a list of records with the following fields:
\begin{itemize}
  \item \verb!page_id_from!: an integer, the identifier of the source article.
  \item \verb!page_title_from!: a string, the title of the source article;
  \item \verb!page_id_to!: an integer, the identifier of the target article;
  \item \verb!page_title_to!: a string, the title of the target article;
\end{itemize}

If a page contains a link to the same page multiple times, this would appear as multiple rows in the \textsc{WikiLinkSnapshots} dataset.
When transforming this data to graph format we eliminate these multiple occurrences, because we are only interested in the fact that the two pages are linked. Wikipedia policies about linking~\cite{wiki:en:Wikipedia:MOS:UL} state that in general a link should appear only once in an article and discourage contributors to put multiple links to the same destination. One clear example is the page \emph{New York City} where, for example, the expression \emph{``United States''} is used to link to the corresponding article only once, at the first occurrence. For these reasons, we do not think it is justified to assign any special meaning to the fact that two page have multiple direct connections between them.

\noindent Figure~\ref{image:process} summarizes the steps followed to produce the \wlg from the Wikipedia dumps with the intermediate datasets produced.

\subsection{Dataset Description}

\begin{table*}[ht]
\scriptsize
\centering
\begin{tabular}{@{}lrrrrrrrrrr@{}}
\toprule
\multirow{2}{*}{\textbf{date}} & \multicolumn{2}{c}{\textbf{de}} & \multicolumn{2}{c}{\textbf{en}} & \multicolumn{2}{c}{\textbf{es}} & \multicolumn{2}{c}{\textbf{fr}} & \multicolumn{2}{c}{\textbf{it}} \\ \cmidrule(r){2-3}\cmidrule(r){4-5}\cmidrule(r){6-7}\cmidrule(r){8-9}\cmidrule(r){10-11}
 & \multicolumn{1}{c}{\textbf{N}} & \multicolumn{1}{c}{\textbf{E}} & \multicolumn{1}{c}{\textbf{N}} & \multicolumn{1}{c}{\textbf{E}} & \multicolumn{1}{c}{\textbf{N}} & \multicolumn{1}{c}{\textbf{E}} & \multicolumn{1}{c}{\textbf{N}} & \multicolumn{1}{c}{\textbf{E}} & \multicolumn{1}{c}{\textbf{N}} & \multicolumn{1}{c}{\textbf{E}} \\ \toprule
2001-03-01 & 0 & 0 & 37 & 31 & 0 & 0 & 0 & 0 & 0 & 0 \\
2002-03-01 & 900 & 1,913 & 27,654 & 223,705 & 1,230 & 2,664 & 55 & 53 & 0 & 0 \\
2003-03-01 & 14,545 & 126,711 & 118,946 & 1,318,655 & 2,786 & 13,988 & 6,694 & 58,027 & 1,036 & 9,695 \\
2004-03-01 & 63,739 & 794,561 & 248,193 & 3,170,614 & 17,075 & 162,219 & 28,798 & 338,108 & 6,466 & 97,721 \\
2005-03-01 & 244,110 & 3,659,389 & 624,287 & 8,505,195 & 43,114 & 457,032 & 96,676 & 1,238,756 & 32,834 & 355,197 \\
2006-03-01 & 474,553 & 7,785,292 & 1,342,642 & 18,847,709 & 112,388 & 1,351,111 & 283,831 & 3,926,485 & 149,935 & 1,434,869 \\
2007-03-01 & 775,104 & 11,946,193 & 2,425,283 & 34,219,970 & 253,569 & 3,327,609 & 555,471 & 7,900,561 & 302,276 & 3,960,767 \\
2008-03-01 & 1,063,222 & 15,598,850 & 3,676,126 & 50,270,571 & 452,333 & 6,292,452 & 1,113,622 & 12,546,302 & 507,465 & 7,239,521 \\
2009-03-01 & 1,335,157 & 19,607,930 & 4,848,297 & 61,318,980 & 762,234 & 9,504,039 & 1,369,619 & 16,546,043 & 693,445 & 10,713,417 \\
2010-03-01 & 1,603,256 & 23,834,140 & 5,937,618 & 71,024,045 & 1,159,567 & 12,844,652 & 1,632,118 & 21,064,666 & 877,089 & 14,120,469 \\
2011-03-01 & 1,879,381 & 28,457,497 & 7,027,853 & 82,944,163 & 1,693,815 & 17,454,997 & 1,890,614 & 25,704,865 & 1,043,648 & 17,496,901 \\
2012-03-01 & 2,163,719 & 33,036,436 & 7,922,426 & 93,924,479 & 1,944,529 & 21,167,388 & 2,137,209 & 30,422,158 & 1,213,961 & 21,069,750 \\
2013-03-01 & 2,461,158 & 37,861,651 & 8,837,308 & 105,052,706 & 2,198,429 & 24,314,571 & 2,369,365 & 34,791,331 & 1,377,144 & 24,694,404 \\
2014-03-01 & 2,712,984 & 42,153,240 & 9,719,211 & 116,317,952 & 2,409,026 & 27,090,659 & 2,594,282 & 39,257,288 & 1,511,827 & 26,821,204 \\
2015-03-01 & 2,933,459 & 46,574,886 & 10,568,011 & 127,653,091 & 2,561,516 & 29,529,035 & 2,809,572 & 43,831,574 & 1,643,387 & 29,867,490 \\
2016-03-01 & 3,155,927 & 50,904,750 & 11,453,255 & 139,194,105 & 2,728,713 & 32,633,513 & 3,037,908 & 48,659,900 & 1,802,952 & 32,521,188 \\
2017-03-01 & 3,372,406 & 55,184,610 & 12,420,400 & 150,743,638 & 2,881,220 & 35,546,330 & 3,239,160 & 53,126,118 & 1,917,410 & 35,158,350 \\
2018-03-01 & 3,588,883 & 59,535,864 & 13,685,337 & 163,380,007 & 3,034,113 & 38,348,163 & 3,443,206 & 57,823,305 & 2,117,022 & 37,814,105 \\ \bottomrule
\end{tabular}
\caption{Number of nodes $N$ and edges $E$ for each graph snapshot of \wlg dataset obtained for the English (en), German (de), Spanish (es),
French (fr), and Italian (it) Wikipedia editions.}
\label{table:dataset01}
\end{table*}

\begin{table*}[ht]
\scriptsize
\centering
\begin{tabular}{@{}lrrrrrrrr@{}}
\toprule
\multirow{2}{*}{\textbf{date}} & \multicolumn{2}{c}{\textbf{nl}} & \multicolumn{2}{c}{\textbf{pl}} & \multicolumn{2}{c}{\textbf{ru}} & \multicolumn{2}{c}{\textbf{sv}} \\ \cmidrule(r){2-3}\cmidrule(r){4-5}\cmidrule(r){6-7}\cmidrule(r){8-9}
 & \multicolumn{1}{c}{N} & \multicolumn{1}{c}{E} & \multicolumn{1}{c}{\textbf{N}} & \multicolumn{1}{c}{\textbf{E}} & \multicolumn{1}{c}{\textbf{N}} & \multicolumn{1}{c}{\textbf{E}} & \multicolumn{1}{c}{\textbf{N}} & \multicolumn{1}{c}{\textbf{E}} \\ \toprule
2001-03-01 & 0 & 0 & 0 & 0 & 0 & 0 & 0 & 0 \\
2002-03-01 & 368 & 728 & 698 & 1,478 & 0 & 0 & 122 & 184 \\
2003-03-01 & 5,182 & 41,875 & 8,799 & 68,720 & 108 & 239 & 6,708 & 33,473 \\
2004-03-01 & 23,059 & 225,429 & 24,356 & 299,583 & 1,600 & 3,927 & 22,218 & 171,486 \\
2005-03-01 & 62,601 & 669,173 & 61,378 & 779,843 & 11,158 & 63,440 & 66,673 & 651,671 \\
2006-03-01 & 169,193 & 1,850,260 & 234,506 & 2,218,720 & 64,359 & 422,903 & 163,988 & 1,605,526 \\
2007-03-01 & 338,354 & 3,746,141 & 395,723 & 4,575,510 & 246,494 & 1,849,540 & 269,599 & 2,627,901 \\
2008-03-01 & 523,985 & 6,037,117 & 546,236 & 7,151,435 & 459,863 & 3,762,487 & 370,569 & 3,746,860 \\
2009-03-01 & 667,311 & 7,900,852 & 690,887 & 9,663,964 & 703,316 & 6,395,215 & 452,132 & 4,841,861 \\
2010-03-01 & 764,277 & 9,467,588 & 822,868 & 11,776,724 & 962,680 & 9,881,672 & 542,900 & 5,856,848 \\
2011-03-01 & 879,062 & 11,120,219 & 953,620 & 13,959,431 & 1,295,284 & 13,955,827 & 712,129 & 6,922,100 \\
2012-03-01 & 1,358,162 & 14,255,313 & 1,091,816 & 15,813,952 & 1,562,821 & 17,882,908 & 800,776 & 7,945,812 \\
2013-03-01 & 1,550,027 & 16,241,260 & 1,208,355 & 17,405,307 & 1,862,035 & 21,724,380 & 1,424,006 & 16,812,447 \\
2014-03-01 & 2,332,477 & 19,940,218 & 1,322,701 & 19,244,972 & 2,098,071 & 25,100,193 & 2,422,972 & 26,497,619 \\
2015-03-01 & 2,424,624 & 21,638,960 & 1,414,645 & 20,838,508 & 2,350,262 & 28,242,878 & 3,218,352 & 33,025,219 \\
2016-03-01 & 2,500,880 & 23,252,874 & 1,513,239 & 22,445,122 & 2,782,155 & 31,467,831 & 4,470,345 & 38,864,469 \\
2017-03-01 & 2,569,547 & 24,691,572 & 1,597,694 & 24,238,529 & 3,094,419 & 34,441,603 & 6,062,996 & 51,975,115 \\
2018-03-01 & 2,626,527 & 25,834,057 & 1,684,606 & 25,901,789 & 3,360,531 & 37,394,229 & 6,131,736 & 52,426,633 \\ \bottomrule
\end{tabular}
\caption{Number of nodes $N$ and edges $E$ for each graph snapshot of \wlg dataset obtained for the Dutch (nl), Polish (pl), Russian (ru),
and Swedish (sv) Wikipedia editions.}
\label{table:dataset02}
\end{table*} 
The \wlg dataset comprises data from $9$ Wikipedia language editions: German (\texttt{de}), English (\texttt{en}), Spanish (\texttt{es}), French (\texttt{fr}), Italian (\texttt{it}), Dutch (\texttt{nl}), Polish (\texttt{pl}), Russian (\texttt{ru}), and Swedish (\texttt{sv}). These editions are the top-$9$ largest editions per number of articles, which also had more than $1,000$ active users~\cite{wiki:en:List_of_Wikipedias}. We excluded Cebuano Wikipedia, because notwithstanding being at the moment the second-largest Wikipedia, its disproportionate growth  with respect to the number of its active users has recently been fueled by massive automatic imports of articles. For fairness, we note that also the growth of Swedish Wikipedia has been led in part by automatic imports of data~\cite{wiki:en:List_of_Wikipedias}, but we have decided to keep it in given it has a reasonably large active user-base.

The \wlg dataset comprises $172$ files for a total of $142$ GB; the average size is $244$~MB and the largest file is $2.4$~GB. For each of the $9$ languages, $18$ files are available with the snapshots of the \emph{wikilink} graph taken on March, 1st from 2001 to 2018. As specified in Section~\ref{sss:graph_snapshots}, these are CSV files that are later compressed in the standard gzip format. The remaining $10$ files contain the hash-sums to verify the integrity of files and a \texttt{README}.

\subsubsection{Where to Find the \wlg Dataset and Its Supporting Material}
The \wlg dataset is published on Zenodo at \url{https://zenodo.org/record/2539424} 
and can be referenced with the DOI number \texttt{10.5281/zenodo.2539424}. 
All other supporing datasets are available at \url{https://cricca.disi.unitn.it/datasets/}. The code used for data processing has been written in Python 3 and it is available on GitHub under the \emph{WikiLinkGraph} 
organization \url{https://github.com/WikiLinkGraphs}.

All the datasets presented in this paper are released under the \emph{Creative Commons - Attribution - 4.0 International} (CC-NY 4.0) license\footnote{\url{https://creativecommons.org/licenses/by/4.0/}}; the code is released under the GNU General Public License version 3 or later\footnote{\url{https://www.gnu.org/licenses/gpl-3.0.en.html}}.

\subsubsection{Basic statistics}

Tables~\ref{table:dataset01} and \ref{table:dataset02} present the number of nodes ($N$) and edges ($E$) for each snapshot included in the \wlg dataset. The number of nodes is much larger that the number of ``content articles'' presented in the main pages of each Wikipedia version. For reference, in March, 2018 English Wikipedia had 5.6M articles~\cite{wikimedia:stats}, however in our snapshot there are more than 13.6M nodes. This is due to the fact that we have left in the graph redirected nodes, as described above, whilst we have resolved the links pointing to them; redirects remain as orphan nodes in the network, receiving no links from other nodes, and having one outgoing link.

\begin{figure}[h!]
\includegraphics[width=0.95\columnwidth]{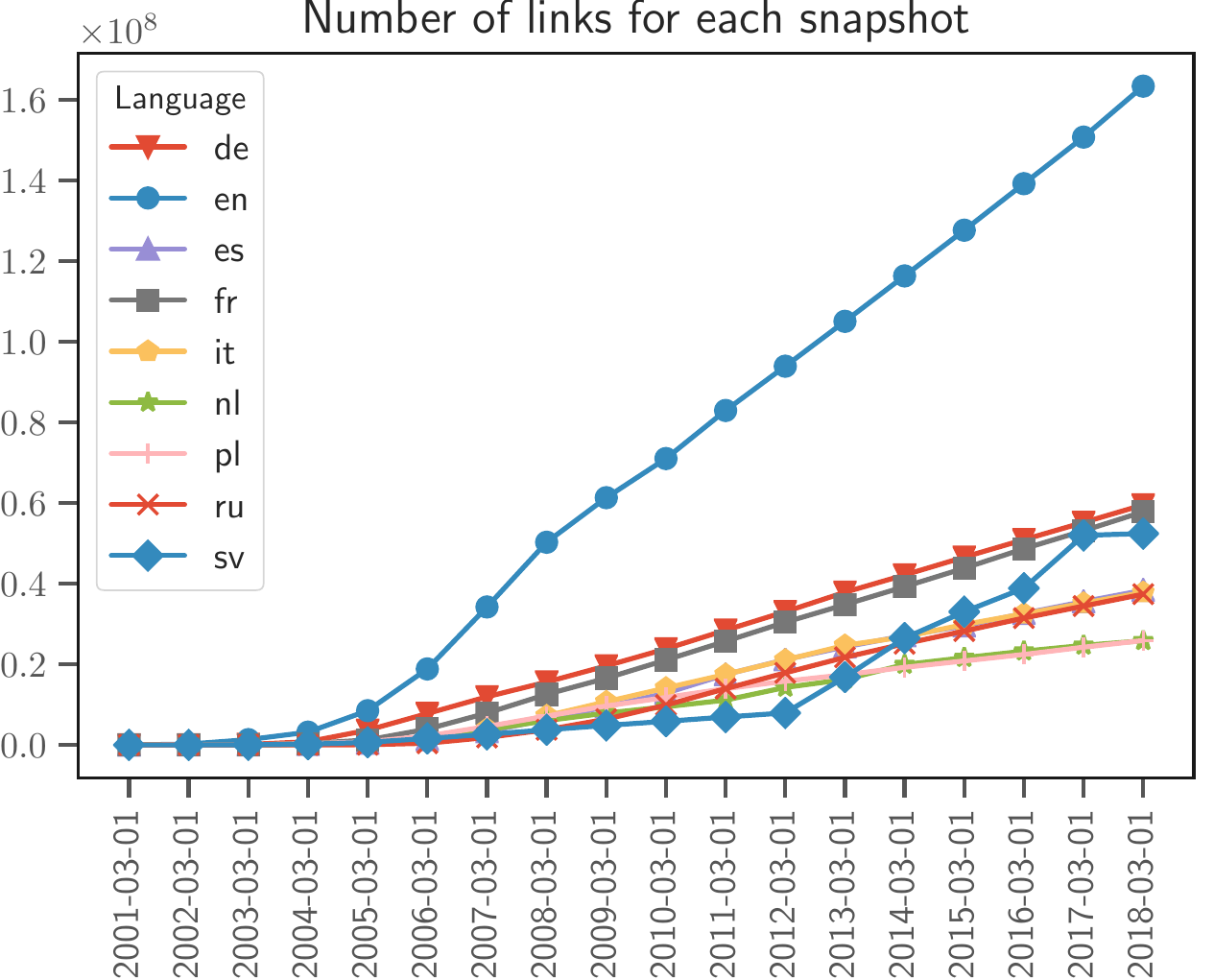}
\caption{\label{image:links} Overview of the growth over time of the number of links in each snapshot in the \wlg dataset.}
\end{figure}

Figure~\ref{image:links} shows a plot of the growth over time of the number of links in the \wlg of each language we have processed. The plot is drawn in linear scale to give a better sense of the relative absolute proportions among the different languages. %
After the first years all language editions exhibit a mostly stable growth pattern with the exception of Swedish, that experienced anomalous growth peaks probably due to massive bot activity.

\section{Analysis and Use Cases}

In this Section we analyse the \wlg dataset to provide some useful insights in the data that will help to demonstrate %
the opportunities opened by this new dataset.

\subsubsection{Comparison with Wikimedia's \textsc{pagelinks} Database Dump.}

To start, we compare our dataset with an existing one provided by the Wikimedia Foundation: the \textsc{pagelinks} table dump.\footnote{For the latest versions of the database dumps, all Wikipedia hyperlinks are available in the "pagelinks" files at \url{https://dumps.wikimedia.org/}.}
 This table tracks all the internal links in a wiki~\cite{mw:Manual:Pagelinks_table}, whether they are links in non-articles pages, link pages across different namespaces, or if they are \emph{transcluded} in a page with a template\footnote{We take the occasion to point out that throughout this paper we refer to "internal links" or \emph{wikilinks} only as links between articles of the encyclopedia, however Wikipedia guidelines use the term more interchangeably to refer both to "links between articles" and "all the links that stay within the project", i.e. including links in other namespaces or that go across namespaces. Whilst it seems that the same confusion exists among the contributors of the encyclopedia, we have decided here to adopt the view for which the proper \emph{wikilinks} are only the links between articles of the encyclopedia.}. The table presents information about the source page identifier and namespace, and the linked-to article title and namespace. There are no duplicates of the same combination of source page id, source page namespace and target title. For this reason, only distinct links in a page are recorded in the table. When updating this table, MediaWiki does not check if the target page exists or not.

\begin{table}[ht]
\resizebox{.95\columnwidth}{!}{
\begin{tabular}{@{}lrrr@{}}
\toprule
\multicolumn{1}{c}{\textbf{lang}} & \multicolumn{1}{c}{\textbf{\textsc{pagelinks} all}} & \multicolumn{1}{c}{\textbf{\textsc{pagelinks} ns0}} & \multicolumn{1}{c}{\textbf{WLG}} \\ \midrule
de	& 156,770,699        & 106,488,110         & 59,535,864         \\
en	& 1,117,233,757      & 476,959,671         & 163,380,007        \\
es	& 88,895,487         & 51,579,346          & 38,348,163         \\
fr	& 270,129,151        & 144,469,298         & 57,823,305         \\
it	& 187,013,995        & 118,435,117         & 37,814,105         \\
nl	& 88,996,775         & 66,606,188          & 25,834,057         \\
pl	& 131,890,972        & 79,809,667          & 25,901,789         \\
ru	& 152,819,755        & 108,919,722         & 37,394,229         \\
sv	& 133,447,975        & 111,129,467         & 52,426,633         \\ \bottomrule
\end{tabular}
}
\caption{\label{table:pagelinks} Comparison of the number of links between articles in the \texttt{ns0} as they result from Wikimedia's \textsc{pagelinks} database table dump (\textsc{pagelinks} ns0) and from the \wlg dataset (WLG). The total number of rows, counting links between other namespaces is given in (\textsc{pagelinks} all).}
\end{table}
 
Table~\ref{table:pagelinks} present a comparison of the number of links extracted from the \textsc{pagelinks} table and the \wlg.

Links in \wlg are much less because links transcluded from templates are not considered. Given the specific research question or application under consideration, it may be more suitable to include or exclude the links that were added to the page by templates; for example, to reconstruct navigational patterns it may be useful not only to consider links from templates, but also links in the navigational interface of MediaWiki.

\begin{table*}[ht!]
\scriptsize
\centering
\resizebox{.95\textwidth}{!}{
\begin{tabular}{@{}llrlrlrlrlr@{}}
\toprule
\multirow{2}{*}{\textbf{\#}} & \multicolumn{2}{c}{\textbf{de}} & \multicolumn{2}{c}{\textbf{en}} & \multicolumn{2}{c}{\textbf{es}} & \multicolumn{2}{c}{\textbf{fr}} & \multicolumn{2}{c}{\textbf{it}} \\ \cmidrule(r){2-3}\cmidrule(r){4-5}\cmidrule(r){6-7}\cmidrule(r){8-9}\cmidrule(r){10-11} 
 & \textbf{article} & \multicolumn{1}{c}{\textbf{\thead{\scriptsize score \\ ${\scriptscriptstyle (\times 10^{-3})}$ }}} & \textbf{article} & \multicolumn{1}{c}{\textbf{\thead{\scriptsize score \\ ${\scriptscriptstyle (\times 10^{-3})}$ }}} & \textbf{article} & \multicolumn{1}{c}{\textbf{\thead{\scriptsize score \\ ${\scriptscriptstyle (\times 10^{-3})}$ }}} & \textbf{article} & \multicolumn{1}{c}{\textbf{\thead{\scriptsize score \\ ${\scriptscriptstyle (\times 10^{-3})}$ }}} & \textbf{article} & \multicolumn{1}{c}{\textbf{\thead{\scriptsize score \\ ${\scriptscriptstyle (\times 10^{-3})}$ }}} \\ \toprule
1 & Vereinigte Staaten & 1.646 & United States & 1.414 & Estados Unidos & 2.301 & France & 2.370 & Stati Uniti d'America & 3.076 \\
2 & Deutschland & 1.391 & World War II & 0.654 & Espa\~na & 2.095 & \'Etats-Unis & 2.217 & Italia & 1.688 \\
3 & Frankreich & 1.020 & United Kingdom & 0.618 & Francia & 1.281 & Paris & 1.228 & Comuni della Francia & 1.303 \\
4 & Zweiter Weltkrieg & 0.969 & Germany & 0.557 & Idioma ingl{\'e}s & 1.073 & Allemagne & 0.977 & Francia & 1.292 \\
5 & Berlin & 0.699 & The New York Times & 0.527 & Argentina & 0.955 & Italie & 0.812 & Germania & 1.257 \\
6 & {\"O}sterreich & 0.697 & Association football & 0.525 & Alemania & 0.909 & Royaume-Uni & 0.773 & Lingua inglese & 1.228 \\
7 & Schweiz & 0.691 & List of sovereign states & 0.523 & Lat{\'i}n & 0.867 & Anglais & 0.764 & Roma & 0.961 \\
8 & Englische Sprache & 0.620 & \makecell[l]{Race and ethnicity \\ in the United States Census} & 0.500 & Animalia & 0.866 & Fran{\c c}ais & 0.748 & Centrocampista & 0.861 \\
9 & Italien & 0.614 & India & 0.491 & M\'exico & 0.853 & Esp\`ece & 0.731 & Europa & 0.805 \\
10 & Latein & 0.599 & Canada & 0.468 & Reino Unido & 0.820 & Canada & 0.710 & 2004 & 0.778 \\ \bottomrule
\end{tabular}
}
\caption{\label{table:pagerank01} Top-10 articles with the highest Pagerank score computed over the most recent snaphost of the \wlg dataset (2018-03-01).}
\end{table*}

\begin{table*}[ht!]
\scriptsize
\centering
\resizebox{.95\textwidth}{!}{
\begin{tabular}{@{}llrlrlrlr@{}}
\toprule %
  \multirow{2}{*}{\textbf{\#}} %
& \multicolumn{2}{c}{\textbf{nl}} %
& \multicolumn{2}{c}{\textbf{pl}} %
& \multicolumn{2}{c}{\textbf{ru}} %
& \multicolumn{2}{c}{\textbf{sv}} %
\\ %
  \cmidrule(r){2-3}\cmidrule(r){4-5}\cmidrule(r){6-7}\cmidrule(r){8-9} %
& \multicolumn{1}{c}{\textbf{article}} %
& \multicolumn{1}{c}{\textbf{\thead{\scriptsize score \\ ${\scriptscriptstyle (\times 10^{-3})}$ }}} %
& \multicolumn{1}{c}{\textbf{article${}^{\ddagger}$}} %
& \multicolumn{1}{c}{\textbf{\thead{\scriptsize score \\ ${\scriptscriptstyle (\times 10^{-3})}$ }}} %
& \multicolumn{1}{c}{\textbf{article${}^{\ddagger}$}} %
& \multicolumn{1}{c}{\textbf{\thead{\scriptsize score \\ ${\scriptscriptstyle (\times 10^{-3})}$ }}} %
& \multicolumn{1}{c}{\textbf{article}} %
& \multicolumn{1}{c}{\textbf{\thead{\scriptsize score \\ ${\scriptscriptstyle (\times 10^{-3})}$ }}} %
\\ %
  \toprule %
1 & Kevers & 3.787 & Stany Zjednoczone & 2.763 & Soedinjonnye Shtaty Ameriki & 3.290 & Familj (biologi) & 5.489 \\
2 & Vlinders & 3.668 & Polska & 2.686 & Sojuz Sovetskih Socialisticheskih Respublik & 2.889 & Sl{\"a}kte & 5.184 \\
3 & Dierenrijk & 3.294 & Francja & 2.360 & Rossija & 2.233 & Nederb{\"o}rd & 4.696 \\
4 & Vliesvleugeligen & 3.084 & Jezyk angielski & 2.110 & Francija & 1.190 & Grad Celsius & 4.144 \\
5 & Insecten & 2.164 & {\L}acina & 1.914 & Moskva & 1.135 & Djur & 4.114 \\
6 & Geslacht (biologie) & 2.101 & Niemcy & 1.698 & Germanija & 1.080 & Catalogue of Life & 3.952 \\
7 & Soort & 1.954 & W{\l}ochy & 1.229 & Sankt-Peterburg & 0.881 & {\r A}rsmedeltemperatur & 3.878 \\
8 & Frankrijk & 1.932 & Wielka Brytania & 1.124 & Ukraina & 0.873 & {\r A}rsnederb{\"o}rd & 3.366 \\
9 & Verenigde Staten & 1.868 & Wie{\'s} & 1.095 & Velikobritanija & 0.811 & V{\"a}xt & 2.810 \\
10 & Familie (biologie) & 1.838 & Warszawa & 1.083 & Italija & 0.763 & Leddjur & 2.641 \\ \bottomrule
\end{tabular}
}
\caption{\label{table:pagerank02} Top-10 articles with the highest Pagerank score computed over the most recent snaphost of the \wlg dataset (2018-03-01). (${}^{\ddag}$) Russian and Polish Wikipedia article titles are transliterated.}
\end{table*} 
In this sense, \wlg provides a new facet of the links in Wikipedia that was not readily available before. These two dataset can be used in conjunction, also taking advantage of the vast amount of metadata available accompanying the \wlg dataset, such as the \textsc{RawWikilinks} and \textsc{ResolvedRedirects} datasets.

\subsection{Cross-language Comparison of Pagerank Scores}

A simple, yet powerful application that can exploit the \wlg dataset is computing the general Pagerank score over the latest snapshot available~\cite{brin1998anatomy}. Pagerank borrows from bibliometrics the fundamental idea that being linked-to is a sign of relevance~\cite{franceschet2010pagerank}. This idea is also valid on Wikipedia, whose guidelines on linking state that:
\begin{quote}
``Appropriate links provide instant pathways to locations within and outside the project that are likely to increase readers' understanding of the topic at hand.''~\cite{wiki:en:Wikipedia:MOS:UL}
\end{quote}
In particular, articles should link to articles with relevant information, for example to explain technical terms.

Tables~\ref{table:pagerank01} and \ref{table:pagerank02} presents the Pagerank scores obtained by running the implementation of the Pagerank algorithm from the \texttt{igraph} library\footnote{\url{https://igraph.org/c/doc/igraph-Structural.html#igraph_pagerank}}.

Across $7$ out of the $9$ languages analysed, the Wikipedia article about the \emph{United States} occupies a prominent position being either the highest or the second-highest ranked article in direct competition with articles about countries were the language is spoken. In general, we see across the board that high scores are gained by articles about countries and cities that are culturally relevant for the language of the Wikipedia edition under consideration.

Remarkably, Dutch and Swedish Wikipedia present very different types of articles in the top-10 positions: they are mainly about the field of biology. A detailed investigation of the results and the causes for these differences is beyond the scope of this paper, but we can hypothesize differences in the guidelines about linking that produce such different outcomes.

\section{Research Opportunities using the WikiLinkGraphs Dataset}

The \wlg dataset and its supporting dataset can be useful for research in a variety of contexts. Without pretending to be exhaustive, we present here a few examples.

\subsubsection{Graph Streaming.}
Stream data processing has gained particular consideration in recent years since it is well-suited for a wide range of applications, and streaming sources of data are commonplace in the big data era~\cite{karimov2018benchmarking}. The \wlg dataset, together with the \textsc{RawWikilinks} dataset, can be represented as a graph stream, i.e. a collection of events such as node and link additions and removals. Whilst other datasets are already available for these kind of problems, such as data from social networks, \wlg, being open, can facilitate the reproducibility of any research in this area and can be used as a benchmark.

\subsubsection{Link Recommendation.}
\citeauthor{west2015mining}~(\citeyear{west2015mining}) have studied the problem of identifying missing links in Wikipedia using web logs. More recently, \citeauthor{wulczyn2016growing}~(\citeyear{wulczyn2016growing}) have demonstrated that it is possible to produce personalized article recommendations to translate Wikipedia articles across language editions. The \wlg dataset could be used in place of the web logs for a similar study on recommending the addition of links in a Wikipedia language edition based on the fact that some links exist between the same articles in other Wikipedia language editions.

\subsubsection{Link Addition and Link Removal.} 
The problem of predicting the appearance of links in time-evolving networks has received significant attention~\cite{lu2011link}; the problem of predicting their disappearance, on the other hand, is less studied. Preusse and collaborators~\cite{preusse2013structural} investigated the structural patterns of the evolution of  links in dynamic knowledge networks. To do so, they adapt some indicators from sociology and identify four classes to indicate growth, decay, stability and instability of links. Starting from these indicators, they identify the underlying reasons for individual additions and removals of knowledge links. Armada et al.~\cite{armada2014semantic} investigated the link-removal prediction problem, which they call the \emph{unlink prediction}.
Representing the ever-evolving nature of Wikipedia links, the \wlg dataset and the \textsc{RawWikilinks} datasets are a natural venue for studying the dynamics of link addition and link removal in graphs. 

\subsubsection{Anomaly Detection.}
A related problem is the identification of spurious links, i.e., links that have been erroneously observed~\cite{guimera2009missing,zeng2012removing}. An example of the application of this approach is the detection of links to spam pages on the Web~\cite{benczur2005spamrank}. Similarly, the disconnection of nodes has been predicted in mobile ad-hoc networks~\cite{de2005disconnection}.

\subsubsection{Controversy mapping.}
Given the encyclopedic nature of Wikipedia, the network of articles represents an emerging map of the connections between the corresponding concepts. Previous work by~\citeauthor{markusson2016contrasting}~(\citeyear{markusson2016contrasting}) has shown how a subportion of this network can be leveraged to investigate public debate around a given topic, observing its framing and boundaries as emerging from the grouping of concepts in the graph. The availability of the \wlg dataset can foster controversy mapping approaches to study any topical subpart of the network, with the advantage of adding a temporal and a cross-cultural dimension.

\subsubsection{Cross-cultural studies.}
More than 300 language editions of Wikipedia have being created since its inception in 2001~\cite{wiki:en:History_of_Wikipedia}, of which 291 are actively maintained. Despite the strict neutral point of view policy which is a pillar of the project~\cite{wiki:en:Wikipedia:5P,wiki:en:Wikipedia:NPOV}, different linguistic communities will unavoidably have a different coverage and different representations for the same topic, putting stronger focus on certain entities, and or certain connections between entities. As an example, the articles about bullfighting in different languages may have a stronger connection to concepts from art, literature, and historical figures, or to concepts such as cruelty and animal rights~\cite{pentzold2017digging}. Likewise, the networks from different language versions give prominence to different influential historical characters~\cite{aragon2012biographical,eom2015interactions}. The \wlg dataset allows to compare the networks of $9$ editions of Wikipedia, which are not only big editions, but have a fairly large base of contributors. In this paper, we have presented a simple comparison across the $9$ languages represented, and we have found an indicator of the prominence of the United States and the local culture almost across the board. Many more research questions could be addressed with the \wlg dataset.

\section{Conclusions}

The dataset we have presented, \wlg, makes available the complete graph of links between Wikipedia articles in the nine largest language editions.

An important aspect is that the dataset contains only links appearing in the text of an article, i.e. links intentionally added by the article editors. While the Wikimedia APIs and dumps provide access to the currently existing wikilinks, such data represent instead all hyperlinks between pages, including links automatically generated by templates. Such links tend to create cliques, introducing noise and altering the structural properties of the network.
We demonstrated that this is not an anecdotal issue and may have strongly affected previous research, as with our method we obtain less than the half of the links contained in the corresponding Wikimedia pagelinks dump.

Another limitation of the Wikimedia dumps is that data are available only for the current version of Wikipedia or for a recent snapshot; the \wlg dataset instead provides complete longitudinal data, allowing for the study of the evolution of the graph over time. We provided both yearly snapshots and the raw dataset containing the complete history of every single link within the encyclopedia.

The \wlg dataset is currently made available for the nine largest Wikipedia language editions, however we plan to extend it to other language editions. As the code of all steps is made available, other researchers can also extend the dataset by including more languages or a finer temporal granularity.   

Beyond the opportunities for future research presented above, we believe that also research in other contexts can benefit from this dataset, such as Semantic Web technologies and knowledge bases, artificial intelligence and natural language processing.

\section{Acknowledgements}
The authors would like to thank Michele Bortolotti and the team of ``Gestione Sistemi'' at the University of Trento for their support with the HPC cluster.

This work has been supported by the European Union's Horizon 2020 research and innovation programme under the \textsc{EU Engineroom} project, with Grant Agreement n\textsuperscript{o} 780643.

\bibliographystyle{aaai}
\bibliography{DP-ConsonniC.1067}

\end{document}